\documentclass[aps,showpacs,twocolumn,floats,superscriptaddress]{revtex4}
\usepackage{hyperref}
\usepackage{graphicx}
\usepackage{color}

\usepackage{amsmath}

\begin{document}
\title{Electrical characterization of chemical and dielectric passivation of InAs nanowires}

\author{Gregory W. Holloway}
\affiliation{Institute for Quantum Computing, University of Waterloo, Waterloo, Ontario, N2L 3G1, Canada}
\affiliation{Department of Physics and Astronomy, University of Waterloo, Waterloo, Ontario, N2L 3G1, Canada} 
\affiliation{Waterloo Institute for Nanotechnology, University of Waterloo, Waterloo, Ontario N2L 3G1, Canada}

\author{Chris M. Haapamaki}
\affiliation{Institute for Quantum Computing, University of Waterloo, Waterloo, Ontario, N2L 3G1, Canada}
\affiliation{Department of Chemistry, University of Waterloo, Waterloo, Ontario, N2L 3G1, Canada}
\affiliation{Department of Engineering Physics, Centre for Emerging Device Technologies, McMaster University, Hamilton, Ontario L8S 4L7, Canada}

\author{Paul Kuyanov}
\affiliation{Department of Engineering Physics, Centre for Emerging Device Technologies, McMaster University, Hamilton, Ontario L8S 4L7, Canada}

\author{Ray R. LaPierre}
\affiliation{Department of Engineering Physics, Centre for Emerging Device Technologies, McMaster University, Hamilton, Ontario L8S 4L7, Canada}

\author{Jonathan Baugh\footnote{corresponding author: baugh@iqc.ca}}
\affiliation{Institute for Quantum Computing, University of Waterloo, Waterloo, Ontario, N2L 3G1, Canada}
\affiliation{Department of Chemistry, University of Waterloo, Waterloo, Ontario, N2L 3G1, Canada}
\affiliation{Waterloo Institute for Nanotechnology, University of Waterloo, Waterloo, Ontario N2L 3G1, Canada}

\begin{abstract}
The native oxide at the surface of III-V nanowires, such as InAs, can be a major source of charge noise and scattering in nanowire-based electronics, particularly for quantum devices operated at low temperatures. Surface passivation provides a means to remove the native oxide and prevent its regrowth. Here, we study the effects of surface passivation and conformal dielectric deposition by measuring electrical conductance through nanowire field effect transistors treated with a variety of surface preparations. By extracting field effect mobility, subthreshold swing, threshold shift with temperature, and the gate hysteresis for each device, we infer the relative effects of the different treatments on the factors influencing transport. It is found that a combination of chemical passivation followed by deposition of an aluminum oxide dielectric shell yields the best results compared to the other treatments, and comparable to untreated nanowires. Finally, it is shown that an entrenched, top-gated device using an optimally treated nanowire can successfully form a stable double quantum dot at low temperatures. The device has excellent electrostatic tunability owing to the conformal dielectric layer and the combination of local top gates and a global back gate. 
\end{abstract}

\maketitle

\section{Introduction}
Semiconducting nanowires offer a promising platform for a number of electronic and optoelectronic applications, including quantum information processing devices such as spin qubits \cite{vandenberg2013,nadjperge2010}, topological qubits \cite{Mourik2012,alicea2011,sau2011}, and on-demand single photon generation \cite{claudon2010,reimer2012}. In particular, realization of spin qubits with fully-electrical control can be achieved in materials with a strong spin-orbit coupling by confining single electrons in electrostatically defined quantum dots \cite{li2013,Flindt2007,baugh2010}. While prototypical devices have demonstrated the fundamentals of this implementation \cite{vandenberg2013,nadjperge2010,fasth2005,pfund2006,schroer2011}, further engineering is desirable to improve the reproducibility (less wire to wire variation), the tunability and stability of the electrostatic potential. Fluctuations in the electrostatic potential are largely due to charge traps located at the nanowire surface or in the native oxide layer \cite{salfi2011,Holloway2013b}. Chemical passivation, in which a layer of atoms or molecules is covalently bonded to the semiconductor surface, is one method to prevent the oxide from forming and to passivate surface states. Sulfur atoms and sulfur-functionalized molecules are effective at passivating III-V surfaces \cite{Petrovykh2003, tajik2012}, but tend to decay over a few days or weeks in ambient conditions, making them impractical as a permanent solution \cite{Suyatin2007}. Another method to decouple charge noise in planar structures is to bury the active layer under buffer layers such that the surface is well separated from the active region \cite{manasreh1997,bennett2007}. This idea has been applied, with some success, to nanowires by growing an epitaxial shell of a larger band-gap III-V material around the nanowire \cite{Tilburg2010,Holloway2013a}. However, the complex growth kinetics of the shell and multiple side facets limit the number of materials that will grow uniformly around the nanowire. Alternatively, one could deposit a dielectric shell around the nanowire using a conformal deposition technique such as atomic layer deposition (ALD) or plasma enhanced chemical vapor deposition (PECVD). In addition to protecting the nanowire surface from oxidation, these dielectric layers have the advantage of high breakdown voltage and a high dielectric constant, so that metal gates can be deposited directly onto the shell. This provides excellent capacitive coupling for devices such as transistors and gate-defined quantum dots. \\

While a dielectric shell appears to be a promising solution to this problem, care must be taken to prepare the nanowire surface prior to deposition of the shell to ensure a low defect interface between the two materials. Growth of the nanowire native oxide prior to deposition of the shell is expected to cause degradation of this interface. Here, we attempt to solve this problem by combining chemical passivation with deposition of a dielectric shell to realize a more stable transistor device. We survey a variety of passivation techniques by fabricating field effect transistors (FETs) using InAs nanowires that have undergone different combinations of chemical passivation and dielectric deposition. Cryogenic transport measurements were performed on each set of FETs to quantify their electronic properties and stability. As a practical test of these surface processing techniques, the most promising set of nanowires were used to fabricate a top-gated nanowire transistor, in which an electrostatically defined double quantum dot was successfully realized. 

\section{Experimental Details}
Undoped InAs nanowires were grown by vapor-liquid-solid growth from gold seed particles in a gas source molecular beam expitaxy (MBE) system. The GaAs (111)B growth substrate was prepared by depositing a 1 nm thick Au film, and then heating it in situ in the MBE to form nanoparticles. For nanowire growth, In atoms were supplied as monomers from an effusion cell, and As$_2$ dimers were supplied from an AsH$_3$ gas cracker operating at 950 $^\circ$C. Nanowire growth was carried out at a substrate temperature of 420 $^\circ$C, an In impingement rate of 0.5 $\mu$m/hr, and a V/III flux ratio of 4. Nanowires typically had a diameter of $\sim$ 20 - 80 nm that was roughly equal to the Au nanoparticle diameter at the top of each nanowire, indicating negligible sidewall deposition. Transmission electron microscopy has shown these nanowires to have minimal stacking faults and a wurtzite crystalline structure \cite{Gupta2013}. Suppression of stacking faults was achieved by growing nanowires at a low growth rate of $\sim$ 0.5 $\mu$m/hr \cite{shtrikman2009}. \\

Chemical passivation and deposition of the dielectric shell were performed on the nanowires while still on the growth substrate, so that all facets of the nanowire were exposed equally. Three surface treatments were implemented: a hydrofluoric acid (HF) dip, octadecanethiol (ODT) passivation \cite{Hang2008,sun2012,petrovykh2009}, and thermal oxide (ThO) growth. In addition to performing each process individually, sequential implementations were carried out. In all cases, the time between subsequent surface treatments and the time prior to deposition of the dielectric was minimized to suppress native oxide regrowth; the average time between the end of treatment and reaching a high vacuum environment in the dielectric deposition chamber was $\sim 3-4$ minutes. \\

The HF dip was a 5 second dip in a buffered oxide etchant (BOE) consisting of a 10:1 mixture of NH$_4$F and HF. Following etching, the substrate was rinsed for 2 minutes in deionized water. The purpose of the HF etch was to remove the native oxide from the nanowire surface. Of course, without a subsequent passivation step, the oxide will quickly regrow when the nanowires are exposed to air. Hence, the HF dip served mostly as a cleaning step prior to other steps. The molecule ODT was used to passivate the nanowire surface since it consists of a long carbon chain connected to a thiol group which readily bonds to InAs. Under appropriate conditions, these molecules will form a self-assembled monolayer on the nanowire surface, passivating it and preventing oxidation for days or weeks \cite{Hang2008,sun2012,petrovykh2009}. The deposition of the self assembled monolayer was achieved by placing the nanowire substrate in a 5 mmol/L solution of ODT in isopropyl alcohol (IPA). Once in solution, the container was sealed with Parafilm and heated to 60 $^\circ$C for 1 hour. Following the deposition, the substrate was rinsed for 30 seconds in clean, room temperature isopropyl alcohol (IPA), and then dried with nitrogen. The effectiveness of this method was confirmed by performing contact angle measurements (using a drop of deionized water) on planar InAs substrates. Samples treated with ODT showed a contact angle of 105 degrees, substantially larger than the contact angle of 55 degrees measured on untreated pieces. The increase in contact angle shows that a hydrophobic surface was created, consistent with a well-formed ODT layer. \\

Growth of a thermal oxide was carried out in a rapid thermal annealing system with the growth substrate seated in a graphite susceptor. The oxidation process consisted of a 30 second ramp to 300 $^\circ$C in nitrogen, followed by a 2 minute period at 300 $^\circ$C in 5 slpm of oxygen, and finally a 5 min ramp down to room temperature, again in nitrogen. The process was calibrated by oxide growth and measurement on a bulk InAs substrate to grow an oxide $\sim2$ nm thick. Following the thermal oxidation process, a representative nanowire was inspected with TEM and showed a uniform oxide with a thickness of 1.8 nm. The motivation for testing this process was to grow a denser, more uniform oxide compared to the native oxide, to possibly reduce the densities of defects and charge traps. Such a thermal oxide has been reported to improve mobilities in nanoribbons \cite{ko2010}. \\

Two different dielectric layers were studied in this work: SiN$_x$ deposited by PECVD and Al$_2$O$_3$ deposited by ALD. Both deposition systems were connected to a load lock, allowing the growth substrates to be quickly placed in vacuum, minimizing oxide growth prior to the deposition. The PECVD was carried out at 330 $^\circ$C in 30 sccm of silane and 900 sccm of nitrogen. A 20 nm thick layer of SiN$_x$ was grown by applying 40 W RF plasma for 65 seconds. ALD took place at 300 $^\circ$C using a trimethyl aluminum precursor and a pulsed 300 W RF plasma. Each deposition cycle grows 1 \AA{} of material, and was carried out for 200 cycles to produce a 20 nm thick film. Following deposition, the substrates were removed from the system and cooled to room temperature prior to further device processing. A TEM image of a nanowire treated with HF, ODT, and covered in an Al$_2$O$_3$ shell is shown in figure 1a. The amorphous Al$_2$O$_3$ shell has a thickness of $\sim$ 20 nm as expected, and is clearly distinct from the crystalline InAs core.\\

Once all desired surface treatments and depositions were performed on a set of nanowires, they were transferred to device substrates by dry deposition. Device substrates consisted of a 300 nm thick layer of thermally grown SiO$_2$ on degenerately doped Si. Deposited nanowires were located relative to pre-patterned alignment markers using SEM, and source drain contacts were written in PMMA resist using electron beam lithography. Patterns were designed to produce contacts that were 1 $\mu$m wide with channel lengths of either 500 nm or 1 $\mu$m. After pattern development, the devices were etched in buffered oxide etchant (BOE) to remove the shell material in the contact area and ensure ohmic contacts. The duration of the BOE step was dependent on the shell material: nanowires with no shell were etched for 5 seconds, those with an Al$_2$O$_3$ shell for 20 seconds, and those with a SiN$_x$ shell for 30 seconds. The etch times were chosen to be $20 \%$ longer than the time necessary to remove the same thickness of dielectric from a planar substrate. This helped ensure that the shell was removed from all sides of the nanowire, and had little effect on the nanowire itself, since InAs shows negligible etching in HF. Following the contact etching, the devices were rinsed in deionized water and transferred to a metal evaporation system in less than 3 minutes from leaving the etching solution (pump down time $\sim 8$ minutes). To remove any oxide that formed during the interim, the devices were exposed to a gentle Ar ion plasma for 10 minutes immediately prior to deposition of 30/50 nm of Ti/Au. We find this ion milling to be crucial to achieving reproducible ohmic contacts. An SEM image of a typical nanowire FET is shown in figure 1b.\\

\begin{figure}[!t]
\includegraphics[width= 8.6cm]{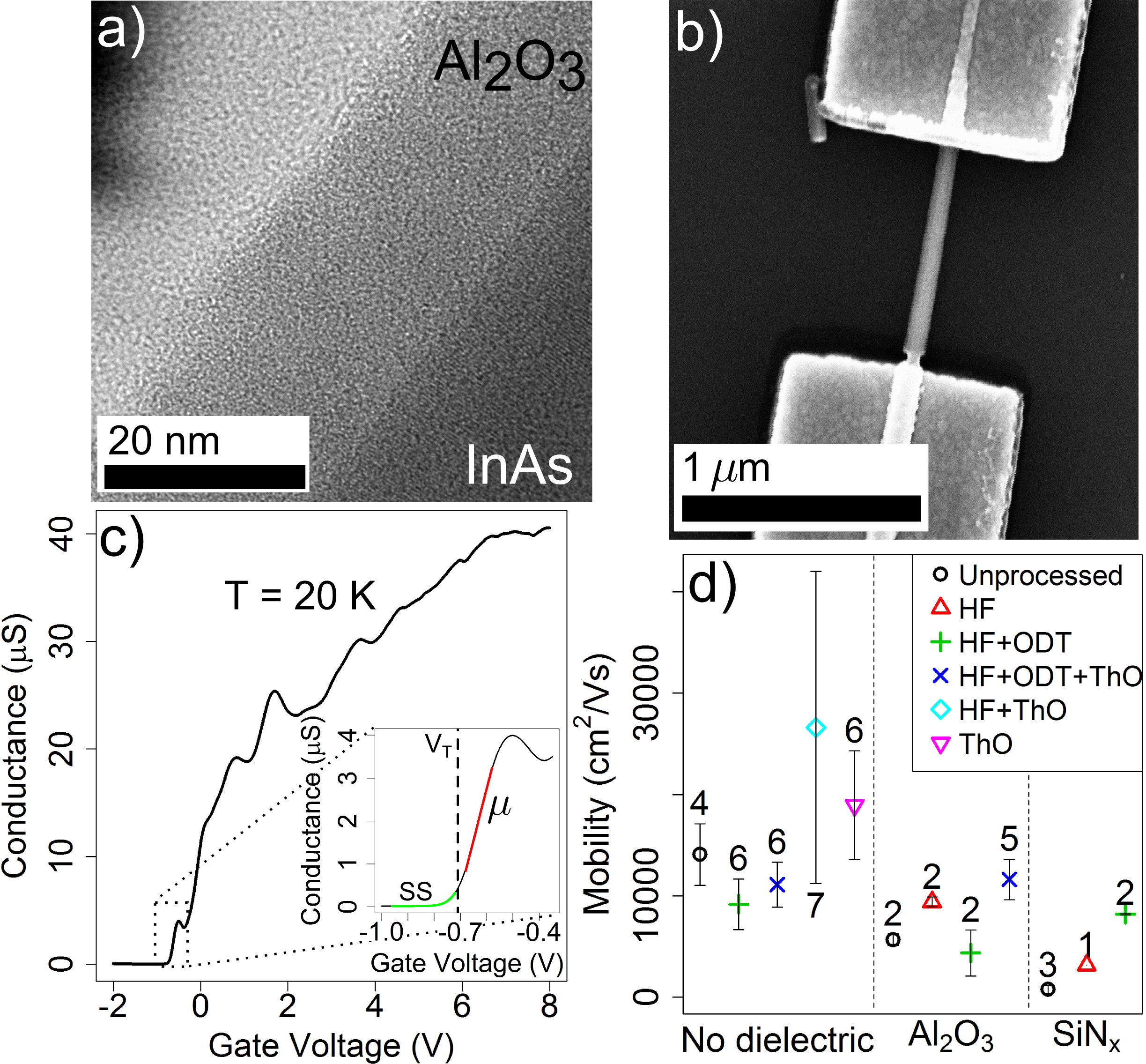} 
\caption{(a) TEM image of a nanowire treated with HF, ODT, and covered in an Al$_2$O$_3$ shell. The amorphous Al$_2$O$_3$ layer is clearly visible around the crystalline InAs core. (b) SEM image of an FET fabricated using a nanowire treated with HF, ODT, and covered in a SiN$_x$ shell. A small section of the shell is unintentionally removed from the channel region near the lower contact, due to lateral etching of the shell material in BOE. (c) Conductance as a function of gate voltage for a representative nanowire FET at 20 K, and a source-drain bias of 1 mV. This was a bare nanowire, with no treatment and no dielectric shell. The inset shows conductance near pinch off, where the black dashed line shows the threshold voltage, and the red and green lines highlight the sections used to extract peak mobility and subthreshold swing, respectively. (d) Average field effect mobility ($\mu_{\text{fe}}$) of nanowire FETs that have undergone different surface processing steps. HF denotes hydrofluoric acid, ODT denotes octadecanthiol, and ThO denotes thermal oxide. The details of each process are described in the main text. Each data point corresponds to the field effect mobility averaged over all devices within a particular surface treatment set, and the number above or below the point indicates the total number of devices in each set. Devices are segregated along the x-axis by their dielectric shell, and the processes prior to shell deposition are indicated by corresponding symbols. Error bars reflect the standard error of the mobility values in each set. } \label{fig1}
\end{figure}

Current-voltage (I-V) measurement of the FETs was carried out in a pumped liquid helium cryostat with a variable temperature controller, allowing temperatures ranging from room temperature to a base of $\sim 1.5$ K. DC electrical characterization was performed in a two-probe configuration using a home-built voltage source and DL Instruments current-voltage preamplifier. FET conductance was modulated by applying a voltage to the degenerately doped silicon substrate that acted as a global back-gate.\\

\section{Results} 
A typical transconductance curve of a FET measured at 20 K is shown in figure 1c (bias voltage $= 1$ mV). The conductance generally increases with applied gate voltage, but shows a few dips which are absent at higher temperatures and are likely due to electron-electron Coulomb interactions and weak localization of charge. To characterize the nanowire channels, field effect mobility was estimated from the transconductance data using the formula \cite{Gupta2013}:

\begin{equation}
\mu_{\text{fe}} = \frac{L^2}{C_g}\frac{dG}{dV_g}
\end{equation}

\noindent where $L$ is the channel length, $G$ is conductance, and $V_g$ is the gate voltage. The gate capacitance, $C_g$, is estimated using a finite element model (COMSOL Multiphysics) of the nanowire FET geometry to include the effects of contact screening. If the contact screening is not taken into account, mobility values for shorter channel FETs are found to be significantly lower than those of similar long channel devices. When contact screening is included, the mobility spread between different channel lengths is effectively removed, allowing for a fair comparison between different FET geometries. The effect of the dielectric shell is neglected in these calculations, as our simulations show it only changes the total gate capacitance by $\leq 3 \%$. Prior to taking the derivative of the conductance, the data was smoothed using a Gaussian moving average with a standard deviation of 10 mV. Since the calculated mobility varies as a function of gate voltage, the highest extracted value is taken as an estimate of the intrinsic mobility. The inset in figure 1c shows conductance near pinch off, where a red line has been fit to the section of the curve with the highest slope, and thus the peak mobility. For this particular device, the peak mobility occurs just above the threshold voltage, but in other devices this is not always the case. This measurement was performed at 20 K, since previous studies have shown that mobility increases at low temperatures, but below 20 K we find that mesoscopic conductance effects become prominent and lead to inaccurate mobility estimates.\\

At low temperature, it has been suggested that mobility in fault-free InAs nanowires is dominated by ionized impurity scattering from surface sites \cite{Gupta2013} or surface roughness \cite{dayeh2010}. A lower value of mobility at 20 K should therefore indicate stronger surface scattering of free electrons. Comparing the mobility estimates across surface processes can therefore give some insight into how each process affects the nanowire surface and/or surface states. Figure 1d shows the average peak mobility across several devices for each process; each point corresponds to a different combination of surface processing steps. FETs are separated along the x-axis by the dielectric shell material, and the symbol for each point denotes the chemical passivation that was carried out prior to shell deposition. One noticeable trend is a decrease in mobility when a dielectric shell is added to the bare nanowire (black circles). This could be due to the formation of defects at the interface between the nanowire's native oxide and the dielectric material. Focusing on nanowires that underwent ODT passivation with no dielectric shell added, we do not see sizeable increases in mobility, as one might expect from removing the native oxide. This could indicate that the ODT layer is ineffective at removing the sources of scattering, the ODT monolayer was incomplete or ill-formed, or it is short-lived such that it decays over the few days between fabrication and measurement. The nanowires with a dielectric shell can provide some insight, since the shell should encase the nanowire and prevent further changes to the nanowire surface. Interestingly, an increase in mobility is generally observed when a surface treatment is performed on the nanowire prior to deposition of the shell. These mobilities never exceed that of the unprocessed bare wires, but are somewhat better than those obtained when depositing dielectric on an unprocessed nanowire. \\

\begin{figure}[t]
\includegraphics[width= 8.6cm]{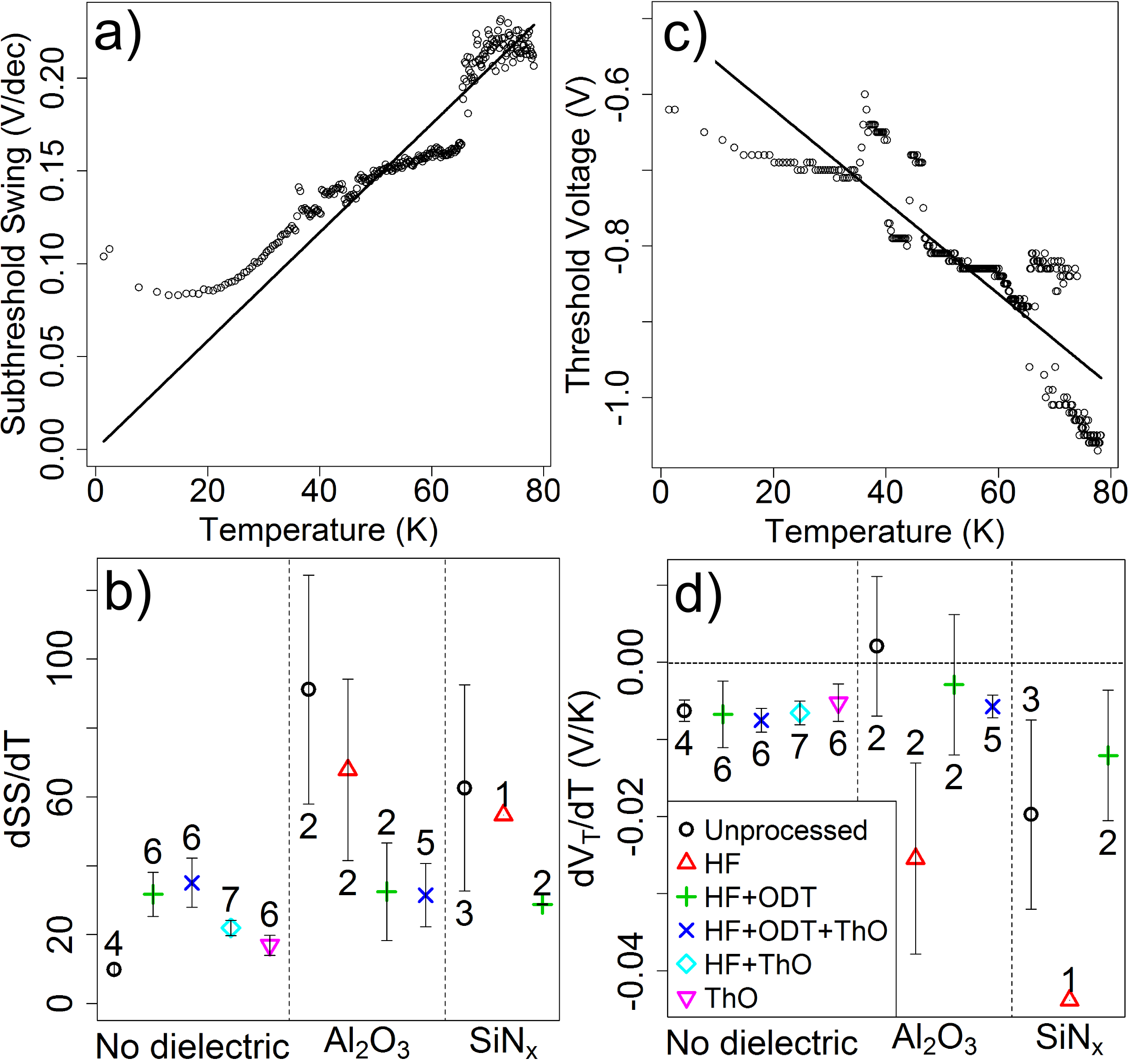} 
\caption{(a,c) Subthreshold swing (SS) and threshold voltage (V$_T$) as a function of temperature ($T$) for an InAs nanowire FET (bare nanowire with no surface processing). Each data set shows a roughly linear temperature dependence, indicated by the linear fit (solid line) in each plot. For (a), this line is a fit to equation 2 to determine $C_{t}$, while for (c) the line is a least squares fit whose slope is related to the density of donor-like states in the nanowire. (b,d) Average dSS/dT and d$V_T$/d$T$ values across different surface processing techniques. Plots have the same layout as figure 1d. For (d), note that most points are located near zero indicating a low density of donor-like states, except for the nanowires treated with HF only, which show a much higher negative value of d$V_T$/d$T$ indicating a higher density of donor-like states. Note that the jumps observed in the data of (a) and (c) are due to slow charge fluctuations in a single trap somewhere near the device channel, i.e. random telegraph noise.} \label{fig2}
\end{figure}

Further understanding of the nanowire/dielectric interface can be gained by looking at the subthreshold swing. The subthreshold swing is a measure of the change in gate voltage needed to drop the current in a transistor by one decade in the region below the threshold voltage. This part of the conductance curve is highlighted in green in the inset of figure 1c. Subthreshold swing was estimated using the formula: $SS = (d\text{log}(I)/dV_g)^{-1}$. Figure 2a shows the value of subthreshold swing as a function of temperature for one device. Above 20 K, the subthreshold swing increases roughly linearly as a function of temperature, as predicted by the following equation \cite{sze2006}:

\begin{equation}
SS = ln(10)\times(k_BT/q)(1+C_{t}/C_g)
\end{equation}

\noindent where $k_B$ is the Boltzmann constant, $T$ is temperature, $q$ is the electron charge, and $C_{t}$ is the capacitance due to interface traps and should be proportional to the density of these traps. The change in subthreshold swing versus temperature can be used as a measure of the relative density of interface traps. The solid line in figure 2a shows the fit of equation 2 to the experimental data. The disagreement below 20 K is likely due to localization and Coulomb blockade effects, which dominate the device conductance at low temperatures and lead to deviations from expected subthreshold swing behavior.  \\

To compare the temperature dependence of the subthreshold swing across different processes, the value of $dSS/dT = (1+C_{t}/C_g)$ is extracted for each device, and the average for each process is plotted in figure 2b. Higher values in this plot correspond to larger values of $C_{t}$, which suggest a larger density of interface traps. The value of $dSS/dT$ for the different processes follows similar trends to the mobility data shown in figure 1d. Most notably, when a dielectric shell is added to a bare nanowire, the density of interface traps increases dramatically. However, by passivating the nanowire surface prior to the dielectric deposition, the density of interface traps can be made comparable to the value seen in wires with no dielectric shell. When looking at  $dSS/dT$ for nanowires with no dielectric shell, the unprocessed devices have the lowest value, followed by the thermal oxide devices. This suggests that the surface processing, particularly HF and ODT treatments, lead to an increase in the density of interface traps. On the other hand, these same processes appear to be essential in reducing the trap densities when the dielectric shell is added. \\

The threshold voltage versus temperature is shown for one device in figure 2c. Threshold voltage is found to decrease roughly linearly as temperature is increased. This is ascribed to thermal activation of donor-like surface states in the nanowire \cite{Dayeh2007}. As temperature increases and more donor-like states become ionized, a more negative gate voltage is required to deplete the nanowire of carriers. The slope of the threshold voltage versus temperature can therefore be used as a rough indication of the surface density of donor-like states. This value is measured by fitting the temperature dependence of the threshold voltage to a linear fit as shown by the solid line in figure 2c. The slopes of these lines are averaged across each process and plotted in figure 2d. Here, the nanowires with no dielectric shell show a similar value of $dV_T/dT$ independent of the surface process. When a dielectric shell is added the values of $dV_T/dT$ are more spread out with no obvious trend apparent. The exception to this is the nanowires treated with only HF prior to the dielectric deposition, which have a much larger value of $dV_T/dT$. This suggests that the HF etch modifies the surface chemistry leaving an increased density of donor-like states, and may be related to hydrogen passivation.\\

\begin{figure}[t]
\includegraphics[width= 8.6cm]{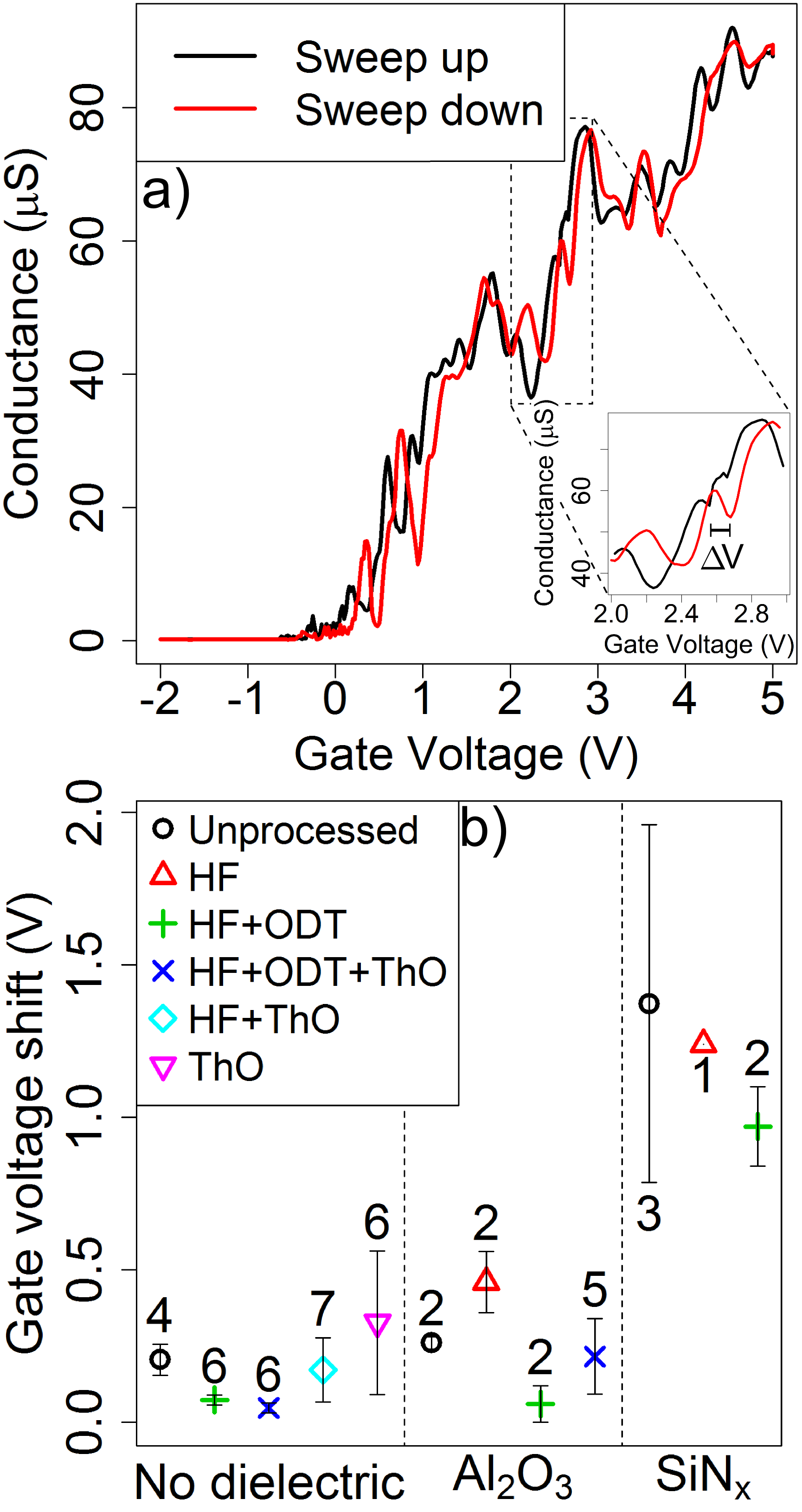} 
\caption{(a) Conductance of a FET at 1.4 K measured by sweeping the gate voltage first up and then down. This particular nanowire was treated with HF, ODT and had a thermal oxide, but no deposited dielectric.  Changes in the trapped charge population cause a hysteretic behavior such that the two curves have a relative gate voltage shift. The inset shows a magnified portion of the conductance curve and the gate voltage shift $\Delta{V}$. (b) Average gate voltage shift at 1.4 K for different surface processes, following the same layout as figure 1d.} \label{fig3}
\end{figure}

The density of charge traps in the FET channel can also be estimated by looking at hysteresis of the conductance curve when sweeping the gate voltage in different directions. This hysteresis is due to the gate voltage modulating the occupation of charge traps \cite{Dayeh2007,Holloway2013b}. Figure 3b shows the hysteresis in FET conductance measured at 1.4 K with a gate voltage sweep rate of 90 mV/s. The inset shows a portion of the conductance curve, where the gate voltage shift $\Delta{V}$ is indicated. The shift is expected to be towards more positive gate voltages when first sweep is up (to positive voltages) and the second sweep is down. This is because just prior to the sweep down, the gate voltage is very positive which fills traps with electrons, and these electrons contribute to an increased negative potential. A more positive gate voltage is then required to cancel out this potential, shifting the conductance curve to more positive values. Figure 3 shows that the data agrees with this expectation. \\

The magnitude of the relative gate voltage shift between the sweep up and sweep down curve, $\Delta{V}$, is an indicator of the density of charge traps. Note that charge traps are not necessarily in the nanowire or its surface, but could be in the dielectric shell or SiO$_2$ substrate, as long as they are close enough to affect the electrostatic potential in the nanowire and thus its conductance. The average gate voltage shifts measured at a temperature of 1.4 K for the different surface processes are shown in figure 3b. Interestingly, it appears that SiN$_x$ has a higher density of traps than either Al$_2$O$_3$ or no dielectric. While the surface passivation techniques may slightly lower the trap density for the nanowires with SiN$_x$, it is still much higher than in the other devices. This suggests that the SiN$_x$ itself likely contains a high density of traps. Conversely, there is not much difference between gate voltage shifts seen in the nanowires with Al$_2$O$_3$ and no dielectric shell, indicating that the Al$_2$O$_3$ shell is not a major source of charge traps. In all cases, the ODT appears to reduce the trap density. Among the devices treated with ODT, those with no dielectric shell and with Al$_2$O$_3$ shell show less hysteresis than untreated nanowires, indicating that ODT could be removing traps at the nanowire surface. \\

\begin{figure}[t]
\includegraphics[width= 8.6cm]{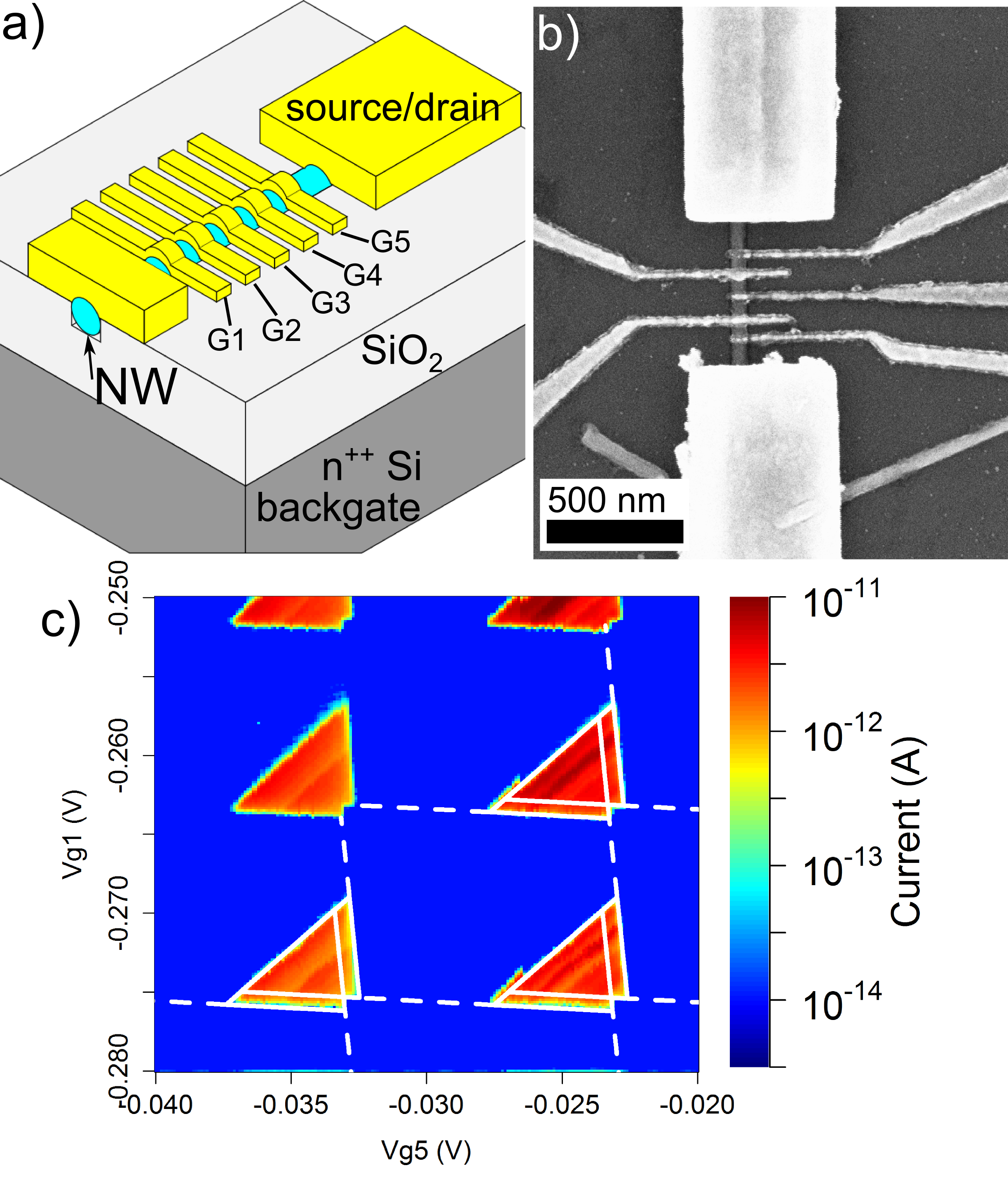} 
\caption{(a) Schematic of the entrenched, top-gated nanowire device for making an electrostatically defined double quantum dot.  The trench geometry allows the top gates to partially wrap around the nanowire while keeping the height profile relatively flat, allowing for fine gates of the desired width (40 nm) and pitch (80 nm). It also allows for global back-gating to tune the carrier density in the nanowire prior to forming barriers and dots. (b) SEM image of a top-gated nanowire FET with dimensions similar to the device studied here. (c) Charge stability diagram, measured at 25 mK and a bias of 2 mV, showing bias triangles that signify double quantum dot transport. The double dot is formed by defining tunnel barriers with gates 1,3, and 5. Here, the electron number in each dot is modulated by sweeping gates 1 and 5. Regions of low current correspond to fixed charge on the two dots. White dashed lines correspond to the best fit of the data to a simple capacitive model \cite{van2002}; the charging energies are 4.5 meV and 4.2 meV for the dots near gate 1 and 5, respectively. The corresponding lever arms for gates 1 and 5 are 0.31 and 0.46, respectively. The back gate voltage was +6 V in this example.} \label{fig4}

\end{figure}

To test the viability of using these nanowires for realizing quantum devices, FETs were fabricated with local top gates which could be used to form an electrostatically defined double quantum dot. Nanowires treated with a combination of ODT and Al$_2$O$_3$ were used for this test, since the results presented above suggested that they showed the best overall characteristics among those devices with a dielectric shell. A schematic of a 5-gated device is shown in figure 4a. In this case, a thinner dielectric shell of only 8 nm was used to increase the capacitive coupling between the nanowire and the local top gates. Nanowires were deposited onto substrates that had been pre-patterned with a series of parallel trenches with a width of 70 nm and a depth of 60 nm, made by reactive ion etching of the SiO$_2$ substrate. Sonicating the substrates for 10 s in acetone following the nanowire deposition increased the number of nanowires in trenches, and simultaneously removed many nanowires not located in trenches (the latter is helpful to prevent stray nanowires from causing breaks or other problems with metal traces connecting the device to bonding pads). Entrenched nanowires were patterned in two electron beam lithography steps to create ohmic source/drain contacts and capacitively coupled top gates, both using Ti/Au metal stacks. Ohmic contacts were made the same way as the previous FETs, using a short BOE dip to remove the dielectric shell. An SEM image of a completed top-gated nanowire FET is shown in figure 4b. \\

Our motivations for using the trench geometry are fourfold: (1) the flatter height profile of the top side of the nanowire with respect to the substrate allows for partial wrap-around gating while maintaining relatively fine widths and pitches (here we achieved 40 nm width and 80 nm pitch). In contrast, we find it impossible to achieve a similar width and pitch when the nanowire is not entrenched, due to a less favourable resist profile during the lithography step. (2) Gating is possible from both top and bottom of the nanowire, in particular we have a global back gate acting from the bottom. This allows for tuning the carrier density to a desired value before using the top gates in depletion mode to form barriers and dots, which in turn allows us to even use nanowires that are normally pinched off at zero back gate voltage. Also, the fact that the nanowire is surrounded on three sides by SiO$_2$ increases the efficiency of the back gate. (3) The nanowire orientation is predefined by the trench alignment, which is very useful when aligning the device to a magnetic field, for example. (4) As mentioned above, sonication allows for the number of stray nanowires not in trenches to be reduced, which is helpful for the device processing. Potential disadvantages of this geometry include a reduced effective surface area for Ohmic contacting, and the possibility for charge noise due to defects in the etched SiO$_2$ sidewalls, however neither appears to be serious in the experimental results we present below. \\

The general strategy for tuning a double quantum dot was to increase the carrier concentration in the nanowire by applying a positive back gate voltage, followed by local depletion due to negative voltages applied to the fine gates to form tunnel barriers. The ability to change the nanowire potential using the global back gate is an important advantage this geometry has over devices which have only local bottom gates \cite{nadjperge2010,fasth2005,pfund2006,schroer2011}, giving better overall electrostatic control. Gates 1, 3 and 5 are used to create a double-well potential. Figure 4c shows the current measured through a device as a function of the voltages on gates 1 and 5, at a lattice temperature of 25 mK and source-drain bias of 2 mV. This shows the characteristic bias triangles and honeycomb structure of a double quantum dot, which can be fit to a simple capacitive model \cite{van2002}. The model fit is indicated by the white dashed line, and from it we extract dot charging energies of 4.5 meV and 4.2 meV. Ideally one would use gates 2 and 4 as plunger gates to control the electron number in each dot, however in this particular device, those gates were weakly coupled to the nanowire due to a defect in the lithography. The lever arm relating gate 1 to its adjacent dot, and gate 5 to its adjacent dot, are 0.31 and 0.46, respectively, indicating very good capacitive coupling. The high dielectric constant of the Al$_2$O$_3$ combined with the partial wrap-around geometry of the gates allows the double quantum dot to be easily formed and tuned with absolute local gate voltages well below 1 V (however in this example, the back gate voltage was +6 V). The data shown in figure 4c was acquired over a time scale of hours, and shows almost no charge noise, reinforcing the results of the FET measurements which suggested this surface treatment would minimize the number of charge traps. Overall, this device geometry along with the ODT and Al$_2$O$_3$ surface processing provide a promising pathway for future mesoscopic devices.\\

\section{Discussion}
Nanowires treated with HF, ODT, and Al$_2$O$_3$ were selected as the most promising candidates for realizing electrostatic quantum dots since they showed a consistently lower density of charge traps than nanowires treated with other surface processes. Despite this favorable charge noise behavior, these nanowires had a lower mobility than most other nanowires that were investigated. One possible explanation for this trend is that mobility is likely dominated by scattering from static scattering sites such as ionized impurities at the surface \cite{Gupta2013}, or surface roughness \cite{dayeh2010}. Importantly, the state of these scatterers need not change with time to affect the mobility, it only matters that they are fixed near the conduction channel to act as scattering sites. This is unlike the dynamic behavior governing the other parameters studied here, where the occupation of charge traps must change to cause the observed effect. For example, gate voltage hysteresis requires traps to change from filled to emptied in order to shift the conductance curve. This suggests that while the HF, ODT, and Al$_2$O$_3$ combination is effective at removing defects which can act as dynamic charge traps, it may induce more static defects which can lead to lower mobility. While the removal of charge noise was deemed more important for realizing electrostatic quantum dot devices, other applications may favor a high mobility and would benefit more from one of the other surface treatments. Our results while limited, suggest that the thermal oxide may improve mobility, similar to previous findings on InAs nanoribbons. \cite{ko2010}\\

One of the motivations for using the dielectric shell was to encase the nanowire so that its electrical properties would remain constant over time and when exposed to ambient conditions. However, we observe that most devices still experience noticeable shifts in threshold voltage after extended periods in a nitrogen atmosphere or air, and this is typically reversible after pumping to a vacuum of $\sim 0.1$ mTorr for several hours. One possible explanation for this is that during the short HF (BOE) etch prior to Ohmic contacting, some lateral etching occurs so that a section of the nanowire channel adjacent to the contacts has its shell removed or partially etched, as shown near the lower contact in figure 1b. This bare region of the channel would be most sensitive to molecular adsorbates. It is also possible that adsorbed molecules on the shell surface transfer charge that affects the nanowire surface potential, however this seems an unlikely mechanism to yield an effect as strong as what is observed. We are exploring alternate techniques for removing the dielectric in the Ohmic regions to prevent this lateral removal of the shell. \\

The molecule chosen for chemical passivation of the nanowire surface was ODT, since it can provide one of the most stable self assembled monolayers on III-V materials. However, it can also be difficult to form a perfect monolayer at the surface, and some reports highlight the necessity of removing all oxygen from the system to achieve ideal passivation \cite{sun2012}. Therefore, it may be more practical to use an ammonium polysulfide process \cite{tajik2012,Suyatin2007} to passivate the surface with S atoms. While this S passivation does not last as long as ODT in ambient conditions, it is stable enough to prevent oxidation during a quick transfer to a dielectric deposition chamber, and is a simpler and potentially more reliable process than ODT passivation. While we observed the ODT step to improve the properties of nanowires that had a subsequent dielectric shell, we did not see noticeable improvements (e.g. in mobility) for nanowires with ODT only as compared to unprocessed nanowires. It would be interesting to see if replacing the ODT step with ammonium polysulfide has a noticeable effect on devices with or without a dielectric shell. \\

\section{Conclusion}
The effects of surface passivation and conformal dielectric deposition on the low temperature electronic properties of InAs nanowire FETs were investigated. It was found that deposition of a dielectric shell on unpassivated nanowires tended to degrade electronic performance, as quantified by mobility, threshold versus temperature, subthreshold swing and gate hysteresis. Al$_2$O$_3$, deposited by an ALD process, was found to be superior to PECVD SiN$_x$. Interestingly, chemical surface passivation prior to dielectric deposition was found to improve electronic performance, in particular nanowires treated with ODT followed by Al$_2$O$_3$ were found to have characteristics similar to unprocessed nanowires. This allows us to maintain the desired intrinsic properties of the nanowire, while encasing it in a conformal insulating high-$k$ dielectric. The addition of this shell facilitated a novel entrenched top-gated device geometry which was used to demonstrate a stable, gate-defined double quantum dot. The dielectric shell improves gate control of the electrostatic potential in the nanowire, evidenced by the strong capacitive coupling between local gates and their adjacent dots. The ability to improve electrostatic control while maintaining intrinsic nanowire transport properties improves the viability of these nanowires as a platform for quantum device applications such as spin and topological qubits. \\

\textbf{Acknowledgements --} We acknowledge the Centre for Emerging Device Technologies at McMaster University and the Quantum NanoFab facility at the University of Waterloo for technical support. Shahram Tavakoli, Nathan Nelson-Fitzpatrick and Roberto Romero provided technical assistance. This work was supported by NSERC, the Canada Foundation for Innovation, the Ontario Centres of Excellence and the Lockheed Martin Corporation.


\begin{thebibliography}{32}
\expandafter\ifx\csname natexlab\endcsname\relax\def\natexlab#1{#1}\fi
\expandafter\ifx\csname bibnamefont\endcsname\relax
  \def\bibnamefont#1{#1}\fi
\expandafter\ifx\csname bibfnamefont\endcsname\relax
  \def\bibfnamefont#1{#1}\fi
\expandafter\ifx\csname citenamefont\endcsname\relax
  \def\citenamefont#1{#1}\fi
\expandafter\ifx\csname url\endcsname\relax
  \def\url#1{\texttt{#1}}\fi
\expandafter\ifx\csname urlprefix\endcsname\relax\def\urlprefix{URL }\fi
\providecommand{\bibinfo}[2]{#2}
\providecommand{\eprint}[2][]{\url{#2}}

\bibitem[{\citenamefont{Van~den Berg et~al.}(2013)\citenamefont{Van~den Berg,
  Nadj-Perge, Pribiag, Plissard, Bakkers, Frolov, and
  Kouwenhoven}}]{vandenberg2013}
\bibinfo{author}{\bibfnamefont{J.}~\bibnamefont{Van~den Berg}},
  \bibinfo{author}{\bibfnamefont{S.}~\bibnamefont{Nadj-Perge}},
  \bibinfo{author}{\bibfnamefont{V.}~\bibnamefont{Pribiag}},
  \bibinfo{author}{\bibfnamefont{S.}~\bibnamefont{Plissard}},
  \bibinfo{author}{\bibfnamefont{E.}~\bibnamefont{Bakkers}},
  \bibinfo{author}{\bibfnamefont{S.}~\bibnamefont{Frolov}}, \bibnamefont{and}
  \bibinfo{author}{\bibfnamefont{L.}~\bibnamefont{Kouwenhoven}},
  \bibinfo{journal}{Physical review letters} \textbf{\bibinfo{volume}{110}},
  \bibinfo{pages}{066806} (\bibinfo{year}{2013}).

\bibitem[{\citenamefont{Nadj-Perge et~al.}(2010)\citenamefont{Nadj-Perge,
  Frolov, Bakkers, and Kouwenhoven}}]{nadjperge2010}
\bibinfo{author}{\bibfnamefont{S.}~\bibnamefont{Nadj-Perge}},
  \bibinfo{author}{\bibfnamefont{S.~M.} \bibnamefont{Frolov}},
  \bibinfo{author}{\bibfnamefont{E.~P. A.~M.} \bibnamefont{Bakkers}},
  \bibnamefont{and} \bibinfo{author}{\bibfnamefont{L.~P.}
  \bibnamefont{Kouwenhoven}}, \bibinfo{journal}{Nature}
  \textbf{\bibinfo{volume}{468}}, \bibinfo{pages}{1084} (\bibinfo{year}{2010}).

\bibitem[{\citenamefont{Mourik et~al.}(2012)\citenamefont{Mourik, Zuo, Frolov,
  Plissard, Bakkers, and Kouwenhoven}}]{Mourik2012}
\bibinfo{author}{\bibfnamefont{V.}~\bibnamefont{Mourik}},
  \bibinfo{author}{\bibfnamefont{K.}~\bibnamefont{Zuo}},
  \bibinfo{author}{\bibfnamefont{S.~M.} \bibnamefont{Frolov}},
  \bibinfo{author}{\bibfnamefont{S.~R.} \bibnamefont{Plissard}},
  \bibinfo{author}{\bibfnamefont{E.~P. A.~M.} \bibnamefont{Bakkers}},
  \bibnamefont{and} \bibinfo{author}{\bibfnamefont{L.~P.}
  \bibnamefont{Kouwenhoven}}, \bibinfo{journal}{Science}
  \textbf{\bibinfo{volume}{336}}, \bibinfo{pages}{1003} (\bibinfo{year}{2012}).

\bibitem[{\citenamefont{Alicea et~al.}(2011)\citenamefont{Alicea, Oreg, Refael,
  von Oppen, and Fisher}}]{alicea2011}
\bibinfo{author}{\bibfnamefont{J.}~\bibnamefont{Alicea}},
  \bibinfo{author}{\bibfnamefont{Y.}~\bibnamefont{Oreg}},
  \bibinfo{author}{\bibfnamefont{G.}~\bibnamefont{Refael}},
  \bibinfo{author}{\bibfnamefont{F.}~\bibnamefont{von Oppen}},
  \bibnamefont{and} \bibinfo{author}{\bibfnamefont{M.~P.}
  \bibnamefont{Fisher}}, \bibinfo{journal}{Nature Physics}
  \textbf{\bibinfo{volume}{7}}, \bibinfo{pages}{412} (\bibinfo{year}{2011}).

\bibitem[{\citenamefont{Sau et~al.}(2011)\citenamefont{Sau, Clarke, and
  Tewari}}]{sau2011}
\bibinfo{author}{\bibfnamefont{J.~D.} \bibnamefont{Sau}},
  \bibinfo{author}{\bibfnamefont{D.~J.} \bibnamefont{Clarke}},
  \bibnamefont{and} \bibinfo{author}{\bibfnamefont{S.}~\bibnamefont{Tewari}},
  \bibinfo{journal}{Physical Review B} \textbf{\bibinfo{volume}{84}},
  \bibinfo{pages}{094505} (\bibinfo{year}{2011}).

\bibitem[{\citenamefont{Claudon et~al.}(2010)\citenamefont{Claudon, Bleuse,
  Malik, Bazin, Jaffrennou, Gregersen, Sauvan, Lalanne, and
  G{\'e}rard}}]{claudon2010}
\bibinfo{author}{\bibfnamefont{J.}~\bibnamefont{Claudon}},
  \bibinfo{author}{\bibfnamefont{J.}~\bibnamefont{Bleuse}},
  \bibinfo{author}{\bibfnamefont{N.~S.} \bibnamefont{Malik}},
  \bibinfo{author}{\bibfnamefont{M.}~\bibnamefont{Bazin}},
  \bibinfo{author}{\bibfnamefont{P.}~\bibnamefont{Jaffrennou}},
  \bibinfo{author}{\bibfnamefont{N.}~\bibnamefont{Gregersen}},
  \bibinfo{author}{\bibfnamefont{C.}~\bibnamefont{Sauvan}},
  \bibinfo{author}{\bibfnamefont{P.}~\bibnamefont{Lalanne}}, \bibnamefont{and}
  \bibinfo{author}{\bibfnamefont{J.-M.} \bibnamefont{G{\'e}rard}},
  \bibinfo{journal}{Nature Photonics} \textbf{\bibinfo{volume}{4}},
  \bibinfo{pages}{174} (\bibinfo{year}{2010}).

\bibitem[{\citenamefont{Reimer et~al.}(2012)\citenamefont{Reimer, Bulgarini,
  Akopian, Hocevar, Bavinck, Verheijen, Bakkers, Kouwenhoven, and
  Zwiller}}]{reimer2012}
\bibinfo{author}{\bibfnamefont{M.~E.} \bibnamefont{Reimer}},
  \bibinfo{author}{\bibfnamefont{G.}~\bibnamefont{Bulgarini}},
  \bibinfo{author}{\bibfnamefont{N.}~\bibnamefont{Akopian}},
  \bibinfo{author}{\bibfnamefont{M.}~\bibnamefont{Hocevar}},
  \bibinfo{author}{\bibfnamefont{M.~B.} \bibnamefont{Bavinck}},
  \bibinfo{author}{\bibfnamefont{M.~A.} \bibnamefont{Verheijen}},
  \bibinfo{author}{\bibfnamefont{E.~P.} \bibnamefont{Bakkers}},
  \bibinfo{author}{\bibfnamefont{L.~P.} \bibnamefont{Kouwenhoven}},
  \bibnamefont{and} \bibinfo{author}{\bibfnamefont{V.}~\bibnamefont{Zwiller}},
  \bibinfo{journal}{Nature communications} \textbf{\bibinfo{volume}{3}},
  \bibinfo{pages}{737} (\bibinfo{year}{2012}).

\bibitem[{\citenamefont{Li et~al.}(2013)\citenamefont{Li, You, Sun, and
  Nori}}]{li2013}
\bibinfo{author}{\bibfnamefont{R.}~\bibnamefont{Li}},
  \bibinfo{author}{\bibfnamefont{J.}~\bibnamefont{You}},
  \bibinfo{author}{\bibfnamefont{C.}~\bibnamefont{Sun}}, \bibnamefont{and}
  \bibinfo{author}{\bibfnamefont{F.}~\bibnamefont{Nori}},
  \bibinfo{journal}{Physical review letters} \textbf{\bibinfo{volume}{111}},
  \bibinfo{pages}{086805} (\bibinfo{year}{2013}).

\bibitem[{\citenamefont{Flindt et~al.}(2007)\citenamefont{Flindt, Sørensen,
  and Flensberg}}]{Flindt2007}
\bibinfo{author}{\bibfnamefont{C.}~\bibnamefont{Flindt}},
  \bibinfo{author}{\bibfnamefont{A.~S.} \bibnamefont{Sørensen}},
  \bibnamefont{and}
  \bibinfo{author}{\bibfnamefont{K.}~\bibnamefont{Flensberg}},
  \bibinfo{journal}{Journal of Physics: Conference Series}
  \textbf{\bibinfo{volume}{61}}, \bibinfo{pages}{302} (\bibinfo{year}{2007}).

\bibitem[{\citenamefont{Baugh et~al.}(2010)\citenamefont{Baugh, Fung, Mracek,
  and LaPierre}}]{baugh2010}
\bibinfo{author}{\bibfnamefont{J.}~\bibnamefont{Baugh}},
  \bibinfo{author}{\bibfnamefont{J.~S.} \bibnamefont{Fung}},
  \bibinfo{author}{\bibfnamefont{J.}~\bibnamefont{Mracek}}, \bibnamefont{and}
  \bibinfo{author}{\bibfnamefont{R.~R.} \bibnamefont{LaPierre}},
  \bibinfo{journal}{Nanotechnology} \textbf{\bibinfo{volume}{21}},
  \bibinfo{pages}{134018} (\bibinfo{year}{2010}).

\bibitem[{\citenamefont{Fasth et~al.}(2005)\citenamefont{Fasth, Fuhrer,
  Bj{\"o}rk, and Samuelson}}]{fasth2005}
\bibinfo{author}{\bibfnamefont{C.}~\bibnamefont{Fasth}},
  \bibinfo{author}{\bibfnamefont{A.}~\bibnamefont{Fuhrer}},
  \bibinfo{author}{\bibfnamefont{M.~T.} \bibnamefont{Bj{\"o}rk}},
  \bibnamefont{and}
  \bibinfo{author}{\bibfnamefont{L.}~\bibnamefont{Samuelson}},
  \bibinfo{journal}{Nano Letters} \textbf{\bibinfo{volume}{5}},
  \bibinfo{pages}{1487} (\bibinfo{year}{2005}).

\bibitem[{\citenamefont{Pfund et~al.}(2006)\citenamefont{Pfund, Shorubalko,
  Leturcq, and Ensslin}}]{pfund2006}
\bibinfo{author}{\bibfnamefont{A.}~\bibnamefont{Pfund}},
  \bibinfo{author}{\bibfnamefont{I.}~\bibnamefont{Shorubalko}},
  \bibinfo{author}{\bibfnamefont{R.}~\bibnamefont{Leturcq}}, \bibnamefont{and}
  \bibinfo{author}{\bibfnamefont{K.}~\bibnamefont{Ensslin}},
  \bibinfo{journal}{Applied physics letters} \textbf{\bibinfo{volume}{89}},
  \bibinfo{pages}{252106} (\bibinfo{year}{2006}).

\bibitem[{\citenamefont{Schroer et~al.}(2011)\citenamefont{Schroer, Petersson,
  Jung, and Petta}}]{schroer2011}
\bibinfo{author}{\bibfnamefont{M.}~\bibnamefont{Schroer}},
  \bibinfo{author}{\bibfnamefont{K.}~\bibnamefont{Petersson}},
  \bibinfo{author}{\bibfnamefont{M.}~\bibnamefont{Jung}}, \bibnamefont{and}
  \bibinfo{author}{\bibfnamefont{J.}~\bibnamefont{Petta}},
  \bibinfo{journal}{Physical review letters} \textbf{\bibinfo{volume}{107}},
  \bibinfo{pages}{176811} (\bibinfo{year}{2011}).

\bibitem[{\citenamefont{Salfi et~al.}(2011)\citenamefont{Salfi, Paradiso,
  Roddaro, Heun, Nair, Savelyev, Blumin, Beltram, and Ruda}}]{salfi2011}
\bibinfo{author}{\bibfnamefont{J.}~\bibnamefont{Salfi}},
  \bibinfo{author}{\bibfnamefont{N.}~\bibnamefont{Paradiso}},
  \bibinfo{author}{\bibfnamefont{S.}~\bibnamefont{Roddaro}},
  \bibinfo{author}{\bibfnamefont{S.}~\bibnamefont{Heun}},
  \bibinfo{author}{\bibfnamefont{S.~V.} \bibnamefont{Nair}},
  \bibinfo{author}{\bibfnamefont{I.~G.} \bibnamefont{Savelyev}},
  \bibinfo{author}{\bibfnamefont{M.}~\bibnamefont{Blumin}},
  \bibinfo{author}{\bibfnamefont{F.}~\bibnamefont{Beltram}}, \bibnamefont{and}
  \bibinfo{author}{\bibfnamefont{H.~E.} \bibnamefont{Ruda}},
  \bibinfo{journal}{ACS nano} \textbf{\bibinfo{volume}{5}},
  \bibinfo{pages}{2191} (\bibinfo{year}{2011}).

\bibitem[{\citenamefont{Holloway
  et~al.}(2013{\natexlab{a}})\citenamefont{Holloway, Song, Haapamaki, LaPierre,
  and Baugh}}]{Holloway2013b}
\bibinfo{author}{\bibfnamefont{G.}~\bibnamefont{Holloway}},
  \bibinfo{author}{\bibfnamefont{Y.}~\bibnamefont{Song}},
  \bibinfo{author}{\bibfnamefont{C.~M.} \bibnamefont{Haapamaki}},
  \bibinfo{author}{\bibfnamefont{R.~R.} \bibnamefont{LaPierre}},
  \bibnamefont{and} \bibinfo{author}{\bibfnamefont{J.}~\bibnamefont{Baugh}},
  \bibinfo{journal}{Journal of Applied Physics} \textbf{\bibinfo{volume}{113}},
  \bibinfo{pages}{024511} (\bibinfo{year}{2013}{\natexlab{a}}).

\bibitem[{Pet(2003)}]{Petrovykh2003}
\bibinfo{journal}{Surface Science} \textbf{\bibinfo{volume}{523}},
  \bibinfo{pages}{231 } (\bibinfo{year}{2003}).

\bibitem[{\citenamefont{Tajik et~al.}(2012)\citenamefont{Tajik, Haapamaki, and
  LaPierre}}]{tajik2012}
\bibinfo{author}{\bibfnamefont{N.}~\bibnamefont{Tajik}},
  \bibinfo{author}{\bibfnamefont{C.}~\bibnamefont{Haapamaki}},
  \bibnamefont{and} \bibinfo{author}{\bibfnamefont{R.}~\bibnamefont{LaPierre}},
  \bibinfo{journal}{Nanotechnology} \textbf{\bibinfo{volume}{23}},
  \bibinfo{pages}{315703} (\bibinfo{year}{2012}).

\bibitem[{\citenamefont{Suyatin et~al.}(2007)\citenamefont{Suyatin, Thelander,
  Björk, Maximov, and Samuelson}}]{Suyatin2007}
\bibinfo{author}{\bibfnamefont{D.~B.} \bibnamefont{Suyatin}},
  \bibinfo{author}{\bibfnamefont{C.}~\bibnamefont{Thelander}},
  \bibinfo{author}{\bibfnamefont{M.~T.} \bibnamefont{Björk}},
  \bibinfo{author}{\bibfnamefont{I.}~\bibnamefont{Maximov}}, \bibnamefont{and}
  \bibinfo{author}{\bibfnamefont{L.}~\bibnamefont{Samuelson}},
  \bibinfo{journal}{Nanotechnology} \textbf{\bibinfo{volume}{18}},
  \bibinfo{pages}{105307} (\bibinfo{year}{2007}).

\bibitem[{\citenamefont{Manasreh}(1997)}]{manasreh1997}
\bibinfo{author}{\bibfnamefont{M.~O.} \bibnamefont{Manasreh}},
  \emph{\bibinfo{title}{Antimonide-related strained-layer heterostructures}},
  vol.~\bibinfo{volume}{3} (\bibinfo{publisher}{CRC Press},
  \bibinfo{year}{1997}).

\bibitem[{\citenamefont{Bennett et~al.}(2007)\citenamefont{Bennett, Boos,
  Ancona, Papanicolaou, Cooke, and Kheyrandish}}]{bennett2007}
\bibinfo{author}{\bibfnamefont{B.~R.} \bibnamefont{Bennett}},
  \bibinfo{author}{\bibfnamefont{J.~B.} \bibnamefont{Boos}},
  \bibinfo{author}{\bibfnamefont{M.~G.} \bibnamefont{Ancona}},
  \bibinfo{author}{\bibfnamefont{N.}~\bibnamefont{Papanicolaou}},
  \bibinfo{author}{\bibfnamefont{G.~A.} \bibnamefont{Cooke}}, \bibnamefont{and}
  \bibinfo{author}{\bibfnamefont{H.}~\bibnamefont{Kheyrandish}},
  \bibinfo{journal}{Journal of electronic materials}
  \textbf{\bibinfo{volume}{36}}, \bibinfo{pages}{99} (\bibinfo{year}{2007}).

\bibitem[{\citenamefont{van Tilburg et~al.}(2010)\citenamefont{van Tilburg,
  Algra, Immink, Verheijen, Bakkers, and Kouwenhoven}}]{Tilburg2010}
\bibinfo{author}{\bibfnamefont{J.~W.~W.} \bibnamefont{van Tilburg}},
  \bibinfo{author}{\bibfnamefont{R.~E.} \bibnamefont{Algra}},
  \bibinfo{author}{\bibfnamefont{W.~G.~G.} \bibnamefont{Immink}},
  \bibinfo{author}{\bibfnamefont{M.}~\bibnamefont{Verheijen}},
  \bibinfo{author}{\bibfnamefont{E.~P. A.~M.} \bibnamefont{Bakkers}},
  \bibnamefont{and} \bibinfo{author}{\bibfnamefont{L.~P.}
  \bibnamefont{Kouwenhoven}}, \bibinfo{journal}{Semiconductor Science and
  Technology} \textbf{\bibinfo{volume}{25}}, \bibinfo{pages}{024011}
  (\bibinfo{year}{2010}).

\bibitem[{\citenamefont{Holloway
  et~al.}(2013{\natexlab{b}})\citenamefont{Holloway, Song, Haapamaki, LaPierre,
  and Baugh}}]{Holloway2013a}
\bibinfo{author}{\bibfnamefont{G.~W.} \bibnamefont{Holloway}},
  \bibinfo{author}{\bibfnamefont{Y.}~\bibnamefont{Song}},
  \bibinfo{author}{\bibfnamefont{C.~M.} \bibnamefont{Haapamaki}},
  \bibinfo{author}{\bibfnamefont{R.~R.} \bibnamefont{LaPierre}},
  \bibnamefont{and} \bibinfo{author}{\bibfnamefont{J.}~\bibnamefont{Baugh}},
  \bibinfo{journal}{Appl. Phys. Lett.} \textbf{\bibinfo{volume}{102}},
  \bibinfo{pages}{043115} (\bibinfo{year}{2013}{\natexlab{b}}).

\bibitem[{\citenamefont{Gupta et~al.}(2013)\citenamefont{Gupta, Song, Holloway,
  Sinha, Haapamaki, LaPierre, and Baugh}}]{Gupta2013}
\bibinfo{author}{\bibfnamefont{N.}~\bibnamefont{Gupta}},
  \bibinfo{author}{\bibfnamefont{Y.}~\bibnamefont{Song}},
  \bibinfo{author}{\bibfnamefont{G.~W.} \bibnamefont{Holloway}},
  \bibinfo{author}{\bibfnamefont{U.}~\bibnamefont{Sinha}},
  \bibinfo{author}{\bibfnamefont{C.~M.} \bibnamefont{Haapamaki}},
  \bibinfo{author}{\bibfnamefont{R.~R.} \bibnamefont{LaPierre}},
  \bibnamefont{and} \bibinfo{author}{\bibfnamefont{J.}~\bibnamefont{Baugh}},
  \bibinfo{journal}{Nanotechnology} \textbf{\bibinfo{volume}{24}},
  \bibinfo{pages}{225202} (\bibinfo{year}{2013}).

\bibitem[{\citenamefont{Shtrikman et~al.}(2009)\citenamefont{Shtrikman,
  Popovitz-Biro, Kretinin, and Heiblum}}]{shtrikman2009}
\bibinfo{author}{\bibfnamefont{H.}~\bibnamefont{Shtrikman}},
  \bibinfo{author}{\bibfnamefont{R.}~\bibnamefont{Popovitz-Biro}},
  \bibinfo{author}{\bibfnamefont{A.}~\bibnamefont{Kretinin}}, \bibnamefont{and}
  \bibinfo{author}{\bibfnamefont{M.}~\bibnamefont{Heiblum}},
  \bibinfo{journal}{Nano Lett.} \textbf{\bibinfo{volume}{9}},
  \bibinfo{pages}{215} (\bibinfo{year}{2009}).

\bibitem[{\citenamefont{Hang et~al.}(2008)\citenamefont{Hang, Wang, Carpenter,
  Zemlyanov, Zakharov, Stach, Buhro, and Janes}}]{Hang2008}
\bibinfo{author}{\bibfnamefont{Q.}~\bibnamefont{Hang}},
  \bibinfo{author}{\bibfnamefont{F.}~\bibnamefont{Wang}},
  \bibinfo{author}{\bibfnamefont{P.~D.} \bibnamefont{Carpenter}},
  \bibinfo{author}{\bibfnamefont{D.}~\bibnamefont{Zemlyanov}},
  \bibinfo{author}{\bibfnamefont{D.}~\bibnamefont{Zakharov}},
  \bibinfo{author}{\bibfnamefont{E.~A.} \bibnamefont{Stach}},
  \bibinfo{author}{\bibfnamefont{W.~E.} \bibnamefont{Buhro}}, \bibnamefont{and}
  \bibinfo{author}{\bibfnamefont{D.~B.} \bibnamefont{Janes}},
  \bibinfo{journal}{Nano Letters} \textbf{\bibinfo{volume}{8}},
  \bibinfo{pages}{49} (\bibinfo{year}{2008}).

\bibitem[{\citenamefont{Sun et~al.}(2012)\citenamefont{Sun, Joyce, Gao, Tan,
  Jagadish, and Ning}}]{sun2012}
\bibinfo{author}{\bibfnamefont{M.~H.} \bibnamefont{Sun}},
  \bibinfo{author}{\bibfnamefont{H.~J.} \bibnamefont{Joyce}},
  \bibinfo{author}{\bibfnamefont{Q.}~\bibnamefont{Gao}},
  \bibinfo{author}{\bibfnamefont{H.~H.} \bibnamefont{Tan}},
  \bibinfo{author}{\bibfnamefont{C.}~\bibnamefont{Jagadish}}, \bibnamefont{and}
  \bibinfo{author}{\bibfnamefont{C.~Z.} \bibnamefont{Ning}},
  \bibinfo{journal}{Nano Lett.} \textbf{\bibinfo{volume}{12}},
  \bibinfo{pages}{3378} (\bibinfo{year}{2012}).

\bibitem[{\citenamefont{Petrovykh et~al.}(2009)\citenamefont{Petrovykh, Smith,
  Clark, Stine, Baker, and Whitman}}]{petrovykh2009}
\bibinfo{author}{\bibfnamefont{D.~Y.} \bibnamefont{Petrovykh}},
  \bibinfo{author}{\bibfnamefont{J.~C.} \bibnamefont{Smith}},
  \bibinfo{author}{\bibfnamefont{T.~D.} \bibnamefont{Clark}},
  \bibinfo{author}{\bibfnamefont{R.}~\bibnamefont{Stine}},
  \bibinfo{author}{\bibfnamefont{L.~A.} \bibnamefont{Baker}}, \bibnamefont{and}
  \bibinfo{author}{\bibfnamefont{L.~J.} \bibnamefont{Whitman}},
  \bibinfo{journal}{Langmuir} \textbf{\bibinfo{volume}{25}},
  \bibinfo{pages}{12185} (\bibinfo{year}{2009}).

\bibitem[{\citenamefont{Ko et~al.}(2010)\citenamefont{Ko, Takei, Kapadia,
  Chuang, Fang, Leu, Ganapathi, Plis, Kim, Chen et~al.}}]{ko2010}
\bibinfo{author}{\bibfnamefont{H.}~\bibnamefont{Ko}},
  \bibinfo{author}{\bibfnamefont{K.}~\bibnamefont{Takei}},
  \bibinfo{author}{\bibfnamefont{R.}~\bibnamefont{Kapadia}},
  \bibinfo{author}{\bibfnamefont{S.}~\bibnamefont{Chuang}},
  \bibinfo{author}{\bibfnamefont{H.}~\bibnamefont{Fang}},
  \bibinfo{author}{\bibfnamefont{P.~W.} \bibnamefont{Leu}},
  \bibinfo{author}{\bibfnamefont{K.}~\bibnamefont{Ganapathi}},
  \bibinfo{author}{\bibfnamefont{E.}~\bibnamefont{Plis}},
  \bibinfo{author}{\bibfnamefont{H.~S.} \bibnamefont{Kim}},
  \bibinfo{author}{\bibfnamefont{S.-Y.} \bibnamefont{Chen}},
  \bibnamefont{et~al.}, \bibinfo{journal}{Nature}
  \textbf{\bibinfo{volume}{468}}, \bibinfo{pages}{286} (\bibinfo{year}{2010}).

\bibitem[{\citenamefont{Dayeh}(2010)}]{dayeh2010}
\bibinfo{author}{\bibfnamefont{S.~A.} \bibnamefont{Dayeh}},
  \bibinfo{journal}{Semiconductor Science and Technology}
  \textbf{\bibinfo{volume}{25}}, \bibinfo{pages}{024004}
  (\bibinfo{year}{2010}).

\bibitem[{\citenamefont{Sze and Ng}(2006)}]{sze2006}
\bibinfo{author}{\bibfnamefont{S.~M.} \bibnamefont{Sze}} \bibnamefont{and}
  \bibinfo{author}{\bibfnamefont{K.~K.} \bibnamefont{Ng}},
  \emph{\bibinfo{title}{Physics of semiconductor devices}}
  (\bibinfo{publisher}{John wiley \& sons}, \bibinfo{year}{2006}).

\bibitem[{\citenamefont{Dayeh et~al.}(2007)\citenamefont{Dayeh, Soci, Yu, Yu,
  and Wang}}]{Dayeh2007}
\bibinfo{author}{\bibfnamefont{S.~A.} \bibnamefont{Dayeh}},
  \bibinfo{author}{\bibfnamefont{C.}~\bibnamefont{Soci}},
  \bibinfo{author}{\bibfnamefont{P.~K.~L.} \bibnamefont{Yu}},
  \bibinfo{author}{\bibfnamefont{E.~T.} \bibnamefont{Yu}}, \bibnamefont{and}
  \bibinfo{author}{\bibfnamefont{D.}~\bibnamefont{Wang}},
  \bibinfo{journal}{Journal of Vacuum Science and Technology B:
  Microelectronics and Nanometer Structures} \textbf{\bibinfo{volume}{25}},
  \bibinfo{pages}{1432} (\bibinfo{year}{2007}).

\bibitem[{\citenamefont{van~der Wiel et~al.}(2002)\citenamefont{van~der Wiel,
  De~Franceschi, Elzerman, Fujisawa, Tarucha, and Kouwenhoven}}]{van2002}
\bibinfo{author}{\bibfnamefont{W.~G.} \bibnamefont{van~der Wiel}},
  \bibinfo{author}{\bibfnamefont{S.}~\bibnamefont{De~Franceschi}},
  \bibinfo{author}{\bibfnamefont{J.~M.} \bibnamefont{Elzerman}},
  \bibinfo{author}{\bibfnamefont{T.}~\bibnamefont{Fujisawa}},
  \bibinfo{author}{\bibfnamefont{S.}~\bibnamefont{Tarucha}}, \bibnamefont{and}
  \bibinfo{author}{\bibfnamefont{L.~P.} \bibnamefont{Kouwenhoven}},
  \bibinfo{journal}{Reviews of Modern Physics} \textbf{\bibinfo{volume}{75}},
  \bibinfo{pages}{1} (\bibinfo{year}{2002}).

\end{thebibliography}
\end{document}